\documentclass[sigconf,authorversion,nonacm]{acmart}

\usepackage{amsmath,amsfonts}
\usepackage{algorithmic}
\usepackage{graphicx}
\usepackage{hyperref}
\usepackage{listings}
\usepackage{textcomp}
\usepackage{siunitx}
\usepackage{multirow}
\usepackage{xcolor}
\usepackage{xspace}
\usepackage{balance}
\usepackage{hyperref}

\def\BibTeX{{\rm B\kern-.05em{\sc i\kern-.025em b}\kern-.08em
    T\kern-.1667em\lower.7ex\hbox{E}\kern-.125emX}}



\usepackage[T1]{fontenc}

\definecolor{backgroundColour}{HTML}{FAFAFA}
\definecolor{keywordclr}{HTML}{3F51B5}
\definecolor{commentclr}{HTML}{757575}
\definecolor{stringsclr}{HTML}{279049}
\definecolor{fnctionclr}{rgb}{0.467, 0, 0.533}
\definecolor{builtinclr}{rgb}{0.35, 0, 0.533}
\definecolor{symbolsclr}{rgb}{0.5, 0.25, 0.25}
\definecolor{numbersclr}{rgb}{0.8, 0.2, 0}
\definecolor{bckgrndclr}{rgb}{0.91, 0.95, 0.95}
\lstdefinestyle{PythonStyle}{
    language=Python,
    backgroundcolor=\color{backgroundColour},
    keywordstyle=\color{keywordclr}\bfseries,
    stringstyle=\color{stringsclr},
    commentstyle=\color{commentclr}\itshape,	basicstyle=\ttfamily\linespread{0.9}\footnotesize,
    breakatwhitespace=false,
    breaklines=true,
    captionpos=b,
    keepspaces=true,
    numbers=left,
    numbersep=5pt,
    numberstyle=\color{commentclr}\ttfamily\tiny,
    showspaces=false,
    showstringspaces=false,
    showtabs=false,
    tabsize=2,
    xleftmargin=2em,
    frame=single,
    framexleftmargin=1.5em,
    morekeywords={assert,with,as}
}
\newcommand{\proxystore}{ProxyStore\xspace}
\newcommand{\globuscompute}{Globus Compute\xspace}
\newcommand{\psendpoint}{PS-endpoint\xspace}
\newcommand{\psendpoints}{PS-endpoints\xspace}
\newcommand{\midway}{Midway2\xspace}

\begin{document}

\title{Accelerating Communications in Federated Applications with Transparent Object Proxies}

\author[J. G. Pauloski]{J. Gregory Pauloski}
\affiliation{
    \institution{University of Chicago}
    \city{}
    \country{}
}
\author[V. Hayot-Sasson]{Valerie Hayot-Sasson}
\affiliation{
    \institution{University of Chicago}
    \city{}
    \country{}
}
\author[L. Ward]{Logan Ward}
\affiliation{
    \institution{Argonne National Laboratory}
    \city{}
    \country{}
}
\author[N. Hudson]{Nathaniel Hudson}
\affiliation{
    \institution{University of Chicago}
    \city{}
    \country{}
}
\author[C. Sabino]{Charlie Sabino}
\affiliation{
    \institution{University of Chicago}
    \city{}
    \country{}
}
\author[M. Baughman]{Matt Baughman}
\affiliation{
    \institution{University of Chicago}
    \city{}
    \country{}
}
\author[K. Chard]{Kyle Chard}
\affiliation{
    \institution{University of Chicago}
    \institution{Argonne National Laboratory}
    \city{}
    \country{}
}
\author[I. Foster]{Ian Foster}
\affiliation{
    \institution{University of Chicago}
    \institution{Argonne National Laboratory}
    \city{}
    \country{}
}

\begin{abstract}
Advances in networks, accelerators, and cloud services encourage programmers to reconsider where to compute---such as when fast networks make it cost-effective to compute on remote accelerators despite added latency.
Workflow and cloud-hosted serverless computing frameworks can manage multi-step computations spanning federated collections of cloud, high-performance computing (HPC), and edge systems, but passing data among computational steps via cloud storage can incur high costs.
Here, we overcome this obstacle with a new programming paradigm that decouples control flow from data flow by extending the pass-by-reference model to distributed applications.
We describe \proxystore{}, a system that implements this paradigm by providing object \emph{proxies} that act as wide-area object references with just-in-time resolution.
This proxy model enables data producers to communicate data unilaterally, transparently, and efficiently to both local and remote consumers.
We demonstrate the benefits of this model with synthetic benchmarks and real-world scientific applications, running across various computing platforms.
\end{abstract}

\keywords{Data Communication and Storage, Distributed Computing, Federated Computing, Open-source Software, Python}

\maketitle

\section{Introduction}

The function-as-a-service (FaaS) and workflow programming para\-digms facilitate the development of scalable distributed applications.
Programmers specify \textit{what} task (e.g., function or workflow stage) to perform without regard to \textit{where} they are executed; the FaaS or workflow system then handles the mechanics of routing each task to a suitable processor.
FaaS systems often assume that tasks are independent, while in workflow systems tasks may be linked in a dependency graph (e.g., a directed acyclic graph). In both cases, it is common for all data movement to pass via a central location such as a FaaS service, workflow engine, shared file system, or task database, where task inputs and outputs can be stored persistently on stable storage.
Such centralized approaches may lead to unnecessary communication~\cite{babuji19parsl,salim2019balsam} but facilitate the implementation of other useful features like re-execution of failed tasks or dynamic adjustments of task location.

The passing of data among tasks via a central location become increasingly problematic when tasks are located on distinct computers.
Consider a program that makes a function call \texttt{x=f()} to produce a value \texttt{x} that is to be consumed by a second function call \texttt{g(x)}.
If \texttt{f()} and \texttt{g()} are dispatched to different computers \emph{C\textsubscript{a}} and \emph{C\textsubscript{b}}, respectively, then \texttt{x} must be transferred from \emph{C\textsubscript{a}} to \emph{C\textsubscript{b}}.
Requiring that this transfer pass via a central location (e.g., FaaS service, workflow controller, shared file system) is inefficient, particularly if \texttt{x} is an intermediary value of no use to the client. Instead, it would be preferable to communicate \texttt{x} directly from \texttt{f()} to \texttt{g()}.
To do this, we need methods for: (1) representing \texttt{x} such that \texttt{f()} and \texttt{g()} can produce and globally reference \texttt{x} and (2) communicating \texttt{x} from \texttt{f()} to \texttt{g()}, despite \texttt{f()} and \texttt{g()} running in different processes, compute nodes, or systems.

\begin{figure}
    \centering
    \includegraphics[width=\columnwidth,trim={0 0px 0 0px},clip]{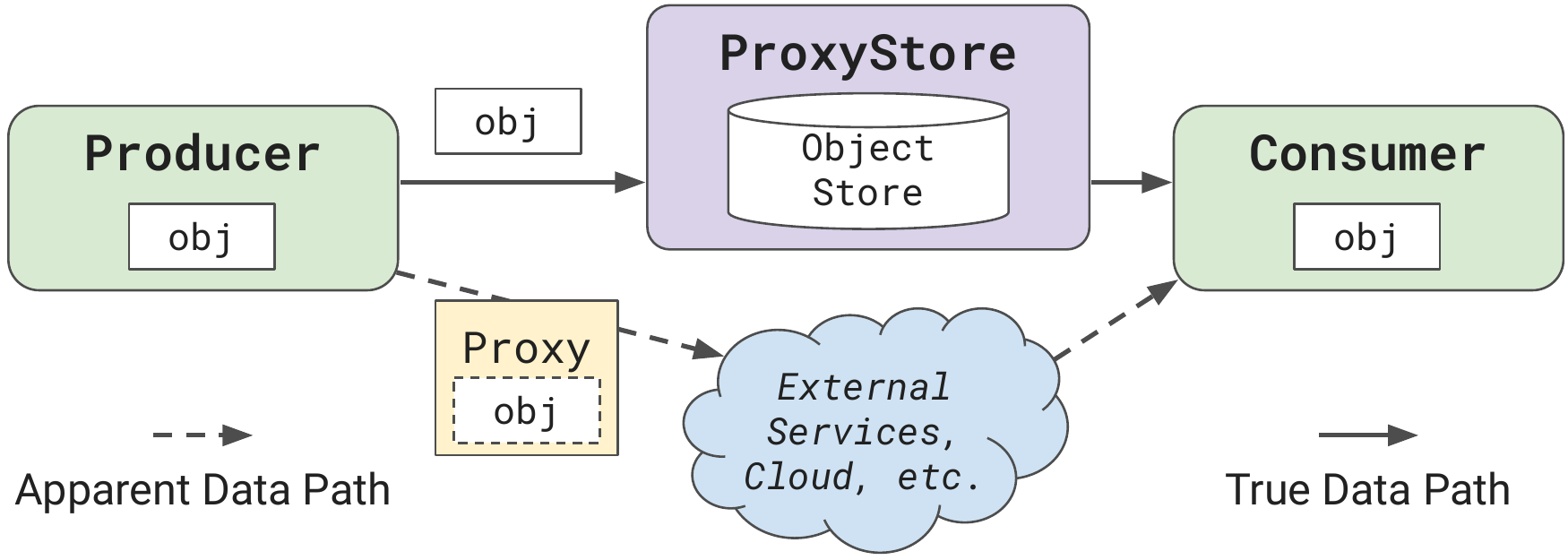}
    \caption{
        \proxystore{} decouples the communication of object data from control flow transparently to the application.
        Data consumers receive lightweight proxies that act like the true object when used, while the heavy lifting of object communication is handled separately.
    }
    \label{fig:proxystore-overview}
\end{figure}

To address these challenges, we present \proxystore{},
an abstraction for managing the routing of data between processes in distributed and federated Python applications.
\proxystore{} allows developers to focus, when writing and composing distributed applications, on logical data flow rather than physical details of where data reside and how data are communicated.
This decoupling enables the dynamic selection of different data movement methods, depending on \emph{what} data are moved, \emph{where} data are moved, or \emph{when} data are moved---a long-standing challenge in distributed application design~\cite{foster1997managing,lumetta1997multi,chiu2002proteus,lastovetsky2006heterompi}.
The \emph{proxy} programming model transparently provides pass-by-reference semantics and just-in-time object resolution to consumers.
A proxy is lightweight and can be communicated efficiently via any means while its referenced object is communicated transparently via optimal routes.
By thus abstracting the use of specialized communication methods, the proxy paradigm improves code compatibility, performance, and productivity.

\proxystore{} provides interfaces to common mediated communication channels (e.g., shared file systems, Globus
\cite{foster2011globus, bryce2012saasglobus, chard2014globus},
Redis~\cite{redis}) and custom implementations that leverage the powerful communication technologies of high-performance computing (HPC) environments and enable direct communication between remote systems.

The contributions of this paper are:
\begin{itemize}
    \item The design and implementation of the proxy model which is, to the best of our knowledge, the first system that transparently provides both pass-by-reference and pass-by-value semantics for distributed applications.
    \item  Data transfer mechanisms that enable fast intra- and inter-site communication in various settings and an extensible framework for seamless integration of new technologies.
    \item Component level benchmarks of \proxystore and comparisons to prior works.
    \item Experiments using \proxystore{} to accelerate real-world federated science applications.
\end{itemize}

The \proxystore{} framework is an open-source Python package available on GitHub and PyPI~\cite{PSgit,PSpypi}.

The rest of this paper is as follows:
\autoref{sec:related} discusses related work in federated and distributed application design;
\autoref{sec:design} outlines \proxystore{} design goals and introduces the core components;
\autoref{sec:implementation} describes the communication channels provided;
\autoref{sec:experiments} presents component-level benchmarks and real-world use cases; and \autoref{sec:conclusion} summarizes our contributions and future plans.

\section{Background and Related Work}
\label{sec:related}
\label{sec:related:communication}

Increasing hardware heterogeneity, faster and more reliable networks, and shifts in application requirements have motivated federated application design, i.e., applications that span several cloud, high-performance (HPC), edge, and personal systems.
Here we discuss technologies that enable the management of computation across diverse systems.

\textbf{Communication decoupling:} The appropriate design for a distributed application's communication fabric depends on the decoupling needed among application processes.
Eugster et al.~\cite{eugster2003pubsub} describe how decoupling can occur along space, time, and synchronization dimensions.
Processes decoupled in \emph{space} interact indirectly via a shared service (e.g., message queue or object store).
A producer and consumer are decoupled in \emph{time} if they need not be active at the same time.
Decoupling in \emph{synchronization} means that data production or consumption does not occur in the primary control flow, so that, for example, processes need not block on communication or can be notified asynchronously of events.

Prior work distinguishes \emph{direct} communication channels from \emph{mediated} channels where ``the communication between participants is done over storage or other indirect means''~\cite{copik2022fmi}.
Direct channels typically provide for rapid communication but prevent space and time decoupling among actors.
Mediated channels necessarily provide space decoupling and can also provide time decoupling if the mediator (e.g., storage) persists for the entirety of the period over which any producers or consumers exist.

\textbf{Data fabrics:}
Tuple spaces, such as in Linda~\cite{linda}, were early shared data fabrics. In the tuple-space model, producers post data as tuples in a shared distributed memory space, from which consumers can retrieve data that match a specified pattern. Tuple spaces have since been implemented
in many languages including Python~\cite{pythontuple}.
DataSpaces~\cite{docan2010dataspaces} provides a tuple-space-like interface to a virtual shared object space designed to support large-scale workflows composed of coupled applications.
The shared space is implemented with the high-speed remote procedure call (RPC) and transfer provided by the Margo and Mercury RPC libraries~\cite{ross2020mochi, sangmin2018argobots, soumagne2013mercury}.
WA-DataSpaces~\cite{aktas2017wa} extends the DataSpaces model to support data staging and predictive prefetching to improve data access times.
The InterPlanetary File System (IPFS) is a decentralized, peer-to-peer file sharing network that provides content-addressing via a flat global namespace~\cite{benet2014ipfs}.

\textbf{Network policies:}
Network access is a core problem in federated computing because policies vary between networks.
Network address translation (NAT) and firewalls often prohibit outside access to local devices, thus preventing direct communication between hosts.
These problems are particularly prevalent in scientific computing where experimental instruments are often in different locations from the data storage and analysis computers.
At some sites, Science DMZs~\cite{data2013sciencedmz} permit bypassing firewalls under programmatic control~\cite{chard2018modern}.
Cross-site data transfer can be performed via cloud services (e.g., in \globuscompute~\cite{chard20funcx}), but this adds latency and can be cost-prohibitive for data-intensive applications.
SciStream~\cite{chung2022scistream} addresses these issues by using gateway nodes (e.g., data transfer nodes in a Science demilitarized zone (DMZ)) to facilitate fast memory-to-memory data transfers between remote hosts.

\textbf{NAT traversal:}
A general solution for communication between two hosts behind separate NATs
is via User Datagram Protocol (UDP) hole punching. In this model, a UDP connection is established between hosts by using a third-party, publicly accessible relay server that facilitates the  connection~\cite{ford2006holepunch}.
For example, Globus Transfer uses such a mechanism for transfers between two Globus Connect Personal endpoints~\cite{chard2014globus}.
The FaaS Messaging Interface (FMI), modeled after the Message Passing Interface (MPI)~\cite{mpi}, provides point-to-point and collective communication for serverless functions~\cite{copik2022fmi}.
It supports both mediated channels, which use external storage accessible by all functions, and direct channels, which use Transmission Control Protocol (TCP) connections that, however, may not be accessible by all function invocations.
When direct TCP communication is not possible, FMI uses a relay server and hole punching to establish a direct connection between function invocations.
Libp2p defines a modular specification for developing peer-to-peer applications with support for NAT traversal~\cite{libp2p,seemann2022decentralized}.
Implementations of the protocol are provided or planned for many popular languages.

\textbf{Workflows:}
Scientific workflow systems (e.g.,
FireWorks~\cite{jain2015fireworks},
Parsl~\cite{babuji19parsl}, Pegasus~\cite{deelman15pegasus},
Radical Pilot~\cite{alsaadi2021radical},
Swift~\cite{wilde11swift}), often include both intra- and inter-site data transfer functionality as a core feature, for movement both of input and output data between clients and execution environments and of intermediate data between tasks~\cite{deelman2009workflows,al2021exaworks}.
Parsl, Pegasus, and Swift all enable transparent intra-site communication via shared file systems, and provide some support for inter-site communication via files.
For example, Parsl supports movement of Python objects via ZeroMQ~\cite{hintjens2013zeromq} sockets in a hub-spoke architecture between the main Parsl process and workers; uni-directional file staging via Hypertext Transfer Protocol (HTTP) and File Transfer Protocol (FTP) (developers must specify URLs for downloading or uploading artifacts); and Globus-based data movement between sites based on user-supplied configuration information specifying the Globus endpoint for each site,
in which case, Parsl inserts data transfer operations in the workflow graph and executes movement before/after task execution.

\textbf{Function-as-a-Service:}
Cloud-hosted serverless frameworks (e.g., Amazon Web Services (AWS) Lambda~\cite{amazonlambda}, Azure Functions~\cite{azurefunctions}, Google Cloud Functions~\cite{googlecloudfunctions}) serialize input and output data along with a function request or result.
Functions can also read and write from cloud object stores (e.g., AWS S3~\cite{aws2021s3}) and pass object IDs as function inputs or outputs.
\globuscompute~\cite{chard20funcx}, formerly \emph{func}X, is a cloud-managed serverless framework that supports remote execution across federated endpoints such as cloud machines, HPC clusters, edge nodes, and workstations.
The \globuscompute cloud service routes each client task to a specified target endpoint and stores results until retrieved by the client.
The cloud service is essential to providing the compute-anywhere features of \globuscompute
but requires that all inputs and results be sent to, and stored in, the cloud (Redis servers hosted in AWS and S3), even if the \globuscompute client and endpoints are located in the same site, which introduces additional latency and costs.
\globuscompute enforces a 5~MB task payload size limit to manage storage and egress costs.

\section{Design and Implementation}
\label{sec:design}

Here we describe the goals and design of the framework, and detail the implementation choices necessary to enable the proxy model.
\proxystore{} provides four primary components: the \texttt{Proxy}, \texttt{Factory}, \texttt{Connector}, and \texttt{Store}.
The \proxystore{} design enables more features and greater flexibility compared to the de facto approaches for mediated communication in federated applications.

\subsection{Assumptions}

We make the following assumptions about usage model and target applications.
(1)~The application requires some combination of space, time, and synchronization decoupling (i.e., \proxystore{} is not intended for highly synchronous applications).
(2)~The application can be described as a composition of dependent tasks that consume and produce Python objects.
We target Python for its pervasiveness in the scientific and workflow systems communities and for the language features that make the proxy model possible.
(3)~Intermediate objects are written only once but may be read many times.
Most task-based workflows fit this paradigm, especially those with pure functional tasks.
(4)~Objects need not be moved to a centralized store, but can stay where they are produced or be moved to where they are to be consumed.
(5)~Users may have their own object storage and communication backends that meet their performance and persistence requirements.
Federated applications that employ FaaS and workflow systems fit these assumptions well.

\subsection{Requirements}
\label{sec:design:requirements}
\proxystore{} must support applications with any of the following attributes:
(1)~data can be produced in many places and must be globally accessible (including across NATs);
(2)~computation can be performed in many places, and regardless of location must be able to consume previously produced data and produce new objects that can then be accessed by others;
(3)~objects may be persistent (must be available for future unknown purposes) or ephemeral (e.g., an intermediate value that is produced by one function and consumed by another, and then never accessed again) and, thus, must exist as long as their associated proxies exist;
(4)~storage locations have varying reliability (e.g., persistent disk vs.\ in-memory) and performance;
(5)~multiple storage or communication methods may need to be employed within a single workload; and
(6)~data consumers need not know the communication method required to access data.

\subsection{The \texttt{Proxy} and \texttt{Factory}}

We meet these design requirements via the use of lazy, transparent object proxies that act as wide-area object references.
The term \textit{proxy} in computer programming refers to an object that acts as the interface for another object.
Proxies are commonly used to add additional functionality to their \emph{target} object or enforce assertions prior to forwarding operations to the target.
For example, a proxy can wrap sensitive objects with access control or provide caching for expensive operations.

Two valuable properties that a proxy can provide are \emph{transparency} and \emph{lazy resolution}.
A \emph{transparent} proxy behaves identically to its target object by forwarding all operations on itself to the target.
For example, given a proxy \texttt{p} of an object \texttt{x}, the types of \texttt{p} and \texttt{x} will be equivalent: i.e., \texttt{isinstance(p, type(x))} and any operation on \texttt{p} will invoke the corresponding operation on \texttt{x}.

A \emph{lazy} or \textit{virtual} proxy provides just-in-time resolution of its target object.
The proxy is initialized with a \emph{factory} rather than the target object.
A factory is an object that is callable like a function and returns the target object.
The proxy is \textit{lazy} in that it does not call the factory to retrieve the target until it is first accessed---a process that is referred to as \emph{resolving} the proxy.
Functionally, proxies have both pass-by-reference and pass-by-value attributes.
The eventual user of the proxied data gets a copy, but unnecessary copies are avoided when the proxy is passed among multiple functions.

This factory-proxy paradigm provides powerful capabilities.
The proxy itself is a lightweight reference to the target that can be communicated cheaply between processes and systems.
The proxy is self-contained because the proxy always contains its factory and the factory includes all logic for data retrieval and manipulation.
That is, the proxy does not need any external information to function correctly.
Proxies eliminate the need for shims or wrapper functions that convert objects into forms expected by downstream code.
Rather, the proxy can be passed to any existing method or function and the conversion is handled internally by the factory.
The consumer code is unaware that the resulting object is anything other than what it expected.
Proxies also have other advantages.
For example: lazy resolution can help amortize costs and avoids unnecessary computation/communication for objects that are never used; nested proxies can enable partial resolution of large objects; and proxies can be moved in place of confidential data (e.g., patient health information) while ensuring that the data can be resolved only where permitted.

\proxystore{} implements lazy transparent object proxies.
The \texttt{Proxy} class implementation is initialized with a factory and intercepts any access to a proxy instance attribute or method; calls the factory to resolve and cache the target object, if the target has not yet been resolved; and forwards the intercepted action to the cached target.
The factory used to initialize a proxy can be any callable Python object (i.e., any object that implements \texttt{\_\_call\_\_}, such as lambdas, functions, and callable class instances).
\texttt{Proxy} modifies its own pickling behavior to include only the factory, not the target, when serializing the proxy, so as to ensure that (1) proxies are small when communicated and (2) a proxy can still be resolved after being communicated to another process.

When a proxy is used, its factory must be able to resolve its target object efficiently.
In a distributed application, this means a factory must be resolvable when the producer and consumer processes exist independently in space or time.
Facilitating this property when processes can exist in the same network or across multiple requires careful consideration for the underlying mediated communication channels used.
We discuss how we achieve this goal with the \texttt{Connector} in the following section.

\subsection{The \texttt{Connector}}

The \texttt{Connector} is a low level interface to a mediated communication channel.
In order to support a wide range of application requirements, we have designed \proxystore{} to be extensible to support various mediated channels that can support different space and time decoupling patterns.
The \texttt{Connector} protocol defines how a client can connect to or operate on a mediated channel, and a \texttt{Connector} implementation must provide four primary operations: \texttt{evict}, \texttt{exist}, \texttt{get}, and \texttt{put}.
The operations act on byte-string data and keys.
E.g., \texttt{put} takes a byte-string to put in the mediated channel and returns a uniquely identifying key (a tuple of metadata); the byte-string is retrievable by calling \texttt{get} on the key.
We chose this model so that third-party code can easily provide new \texttt{Connectors} that are plug-and-play with the rest of \proxystore{}'s features.
A \texttt{Connector} implementation can be either an interface to an external mediated channel (e.g., a Redis server) or a mediated channel itself.
\proxystore provides many \texttt{Connector} implementations that fit both of these categories which we describe further in \autoref{sec:implementation}.

\subsection{The \texttt{Store}}

\begin{figure}
    \centering
    \includegraphics[width=\columnwidth,trim={0 0px 0 0px},clip]{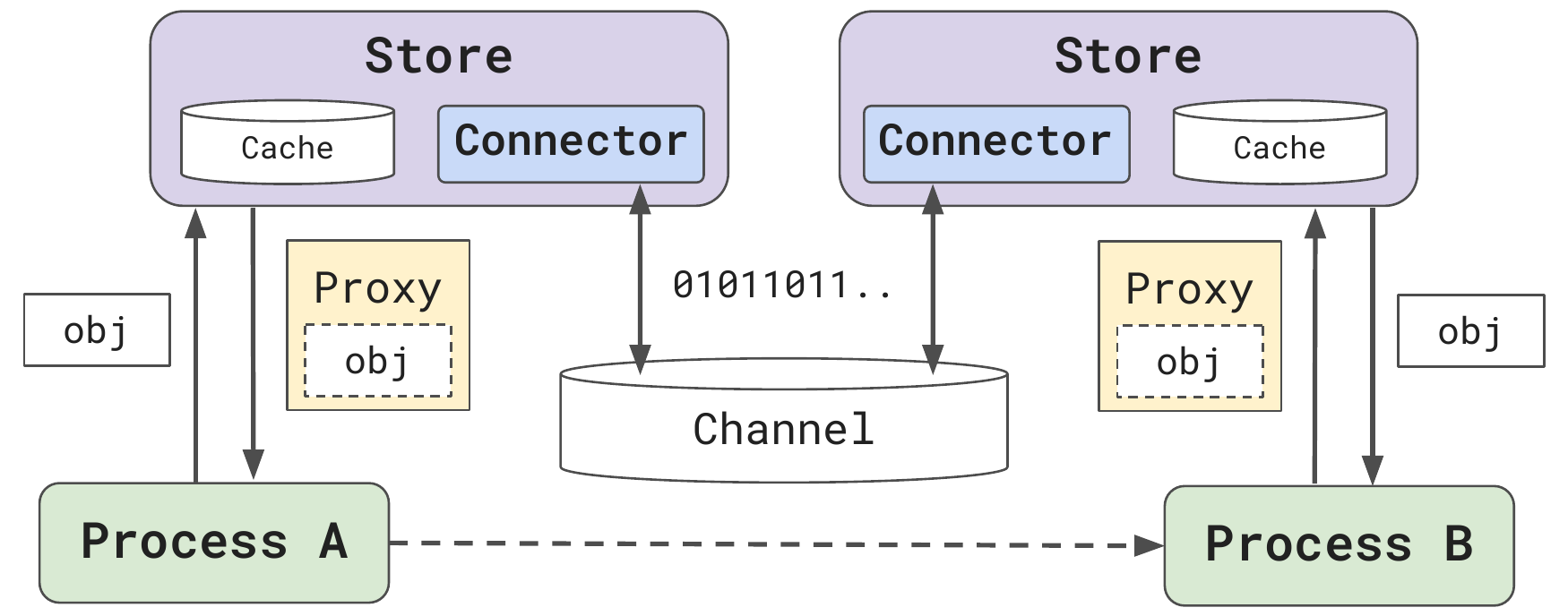}
    \caption{
        Processes interact with a \texttt{Store} to proxy objects, and proxy consuming processes will transparently interact with the local \texttt{Store} instance. The underlying communication is executed using the \texttt{Connector} interface.
    }
    \label{fig:proxystore-schematic}
\end{figure}

\lstset{escapeinside={(*}{*)}}
\begin{lstlisting}[style=PythonStyle, label={lst:proxystore-example}, caption={Example of \proxystore{} usage.}, float, floatplacement=bt]
from (*\MakeLowercase{\proxystore{}}*).connectors.redis import RedisConnector
from (*\MakeLowercase{\proxystore{}}*).proxy import Proxy
from (*\MakeLowercase{\proxystore{}}*).store import Store

def my_function(x: MyDataType) -> ...:
    # x is resolved from "my-store" on first use
    assert isinstance(x, MyDataType)
    # More computation...

store = Store('my-store', RedisConnector(...))

# Store the object and get a proxy
my_object = MyDataType(...)
p = store.proxy(my_object)
assert isinstance(p, Proxy)

my_function(p)  # Succeeds
\end{lstlisting}

\noindent
The \texttt{Store} is the high-level interface used by applications to interact with \proxystore{} as shown in~\autoref{fig:proxystore-schematic}.
A \texttt{Store} is initialized with a \texttt{Connector} instance (a dependency injection pattern) and provides additional functionality on top of the \texttt{Connector}.
Similar to the \texttt{Connector}, the \texttt{Store} exposes \texttt{evict}, \texttt{exist}, \texttt{get}, and \texttt{put} operations; however, these operations act on Python objects rather than byte strings.
The \texttt{Store} (de)serializes objects before invoking the corresponding operation on the \texttt{Connector};
custom (de)serialize functions can be registered with the \texttt{Store} if needed.
The \texttt{Store} also provides caching of operations to reduce communication costs, with caching performed after deserialization to avoid duplicate deserializations.

However, rather than the application invoking the aforementioned operations directly, the \texttt{proxy} method, also provided by the \texttt{Store}, is used.
Calling \texttt{Store.proxy} puts an object in the mediated channel via the \texttt{Connector} instance and returns a proxy (\autoref{lst:proxystore-example}).
The object is serialized before being put in the mediated channel; a factory is generated, containing the key returned by the \texttt{Connector} and additional information necessary to retrieve the object from the mediated channel; and then a new proxy, internalized with the factory, is returned.

An evict flag can be passed when creating a proxy.
If set, the proxy will evict the object from the mediated channel when first resolved.
Subsequently, the \texttt{proxy} operation, alone, is a complete interface to an object store because the \texttt{proxy} method handles the \texttt{put} operation and the proxy resolution process handles \texttt{get}/\texttt{evict}.

The \texttt{Proxy} and \texttt{Factory} instances created by a \texttt{Store} provide functionality for asynchronously resolving the target object in a background thread using the \texttt{resolve\_async} function.
This is useful in code which expects a proxy and wants to overlap the communication of the proxy resolution with other computations.

\texttt{Store} instances are registered globally within a process by name so that initialization is performed only once, caches are shared, and stateful connection objects in the \texttt{Connector} are reused.
Consider a \texttt{Connector} instance $C$ and corresponding \texttt{Store} $S$.
$S$ has been registered in process $P_a$ with name ``my-store'' and is used to create a proxy $p$.
If $p$ is resolved on a remote process $P_b$ where a \texttt{Store} with name ``my-store'' has not yet been registered, $p$ will initialize and register a new \texttt{Store} instance named ``my-store'' with the appropriate \texttt{Connector} when $p$ is resolved.
This is possible because $p$'s factory, created in process $P_a$, includes the appropriate metadata necessary to recreate $C$ and $S$ in process $P_b$.
Subsequent proxies created by any \texttt{Store} with the same name and resolved in $P_b$ will then use the registered \texttt{Store} rather than initializing a new one.

\section{Supported Connectors}
\label{sec:implementation}

\begin{table}
\caption{Summary of provided \texttt{Connector} implementations.}
\label{tab:connectors}
\begin{footnotesize}
\begin{tabular}{llccc}
\toprule
\textbf{Connector} & \textbf{Storage} & \textbf{Intra-Site} & \textbf{Inter-Site} & \textbf{Persistence} \\
\midrule
File      & Disk    & \checkmark &            & \checkmark  \\
Redis     & Hybrid  & \checkmark &            & \checkmark  \\
Margo     & Memory  & \checkmark &            &             \\
UCX       & Memory  & \checkmark &            &             \\
ZMQ       & Memory  & \checkmark &            &             \\
Globus    & Disk    &            & \checkmark & \checkmark  \\
Endpoint  & Hybrid  & \checkmark & \checkmark & \checkmark  \\
\bottomrule
\end{tabular}
\end{footnotesize}
\end{table}

All \texttt{Connector} implementations are built on mediated, byte-level data storage.
Data storage methods are broadly classified as in-memory or on-disk.
Mediated channels use one or both methods, depending on performance and persistence aims.
The proxy abstraction provided by the \texttt{Store} enables a producer to unilaterally (i.e., without the agreement of the receiver) choose the best mediated channel for object communication.

Data storage may be local to the process or machine, within the same network, or at a remote site.
Here, we describe the various \texttt{Connector} implementations provided out-of-the-box that can be used with the \texttt{Store} that support in-memory and on-disk data storage within and between sites (summarized in \autoref{tab:connectors}).
We also describe an implementation provided \texttt{MultiConnector} abstraction which enables intelligent routing of objects across connectors.

\subsection{Intra-Site Communication}
\label{sec:design:intrasite}
Various technologies, such as shared file systems, TCP/UDP sockets, and remote distributed
memory access (RDMA), enable data transfers between nodes on the same local area network: i.e., not located behind different NATs.

\subsubsection{On-disk Storage}
For large objects or data that needs to be persisted, \proxystore{} provides the \texttt{FileConnector} for mediated communication via a shared file system.
The \texttt{FileConnector} is initialized with a path to a data directory in which proxied objects can be serialized and written (and then read) as files.

\subsubsection{Hybrid Storage}

The \texttt{RedisConnector} uses an existing Redis~\cite{redis} or KeyDB~\cite{keydb} server as the mediator.
Redis provides a hybrid between in-memory and on-disk data storage with low-latency, easy configuration, persistence, and optional resilience via replication across nodes.
The \texttt{RedisConnector} implementation is only 31 lines of Python code, exemplifying the ease with which the proxy model can be extended to other mediated communication methods via the \texttt{Connector} protocol.

\subsubsection{Distributed In-memory Storage}
\label{sec:design:intrasite:dim}

Distributed memory backends for intra-site communication permit applications to benefit from increased memory capacity and scalability.
Two implementations are provided, \texttt{MargoConnector} and \texttt{UCXConnector}, to leverage rapid communication on high-speed networks by using the Py-Mochi-Margo~\cite{py-mochi-margo} and UCX-Py~\cite{UCX-Py} libraries, respectively.
A third implementation, \texttt{ZMQConnector}, uses ZeroMQ for communication and is provided as a fallback for compatibility.
When one of these connectors is initialized for the first time in a process, it spawns a process that acts as the storage server for that node.
Thus, these connectors act as interfaces to these spawned servers which make up the actual distributed in-memory store.
These distributed storage methods are elastic---expanding as proxies are propagated to new nodes---and enable the use of state-of-the-art direct communication methods in a mediated fashion.

\subsection{Inter-Site Communication}
\label{sec:design:intersite}

\proxystore{} enables data transfer between computers at different sites (and also between computers at the same site that are located behind different NATs) by using disk-to-disk solutions for bulk data and memory-to-memory solutions for low latency.

\subsubsection{On-disk Storage}

Bulk file transfers between sites are ubiquitous in scientific applications.
To support such transfers, the \texttt{FileConnector} is extended as the \texttt{GlobusConnector} to use Globus to move object files between sites.
Globus transfer supports efficient, secure, and reliable file movement and is widely adopted across computing centers with more than \num{20000} active endpoints.
Globus Connect software is easily deployed on computers without an existing endpoint.

The \texttt{GlobusConnector} is initialized with a mapping of hostname regular expressions to a tuple of \texttt{(Globus Endpoint UUID, Endpoint path)}.
A proxy, while resolving itself, will match the hostname of the current system to the provided hostname regular expressions to determine the directory on the local endpoint with the transferred files.
\texttt{GlobusConnector} keys are the tuple \texttt{(object\_id, task\_id)} where the \texttt{task\_id} is the Globus transfer task ID.
A proxy will wait for the transfer task to succeed before resolving itself or raise an error if there is a Globus transfer failure.

For efficient movement of many objects, the \texttt{Store} provides a \texttt{proxy\_batch} method that will invoke a batch transfer of proxied objects as a single Globus transfer.

\subsubsection{In-memory Storage}

\begin{figure}
    \centering
    \includegraphics[width=\columnwidth]{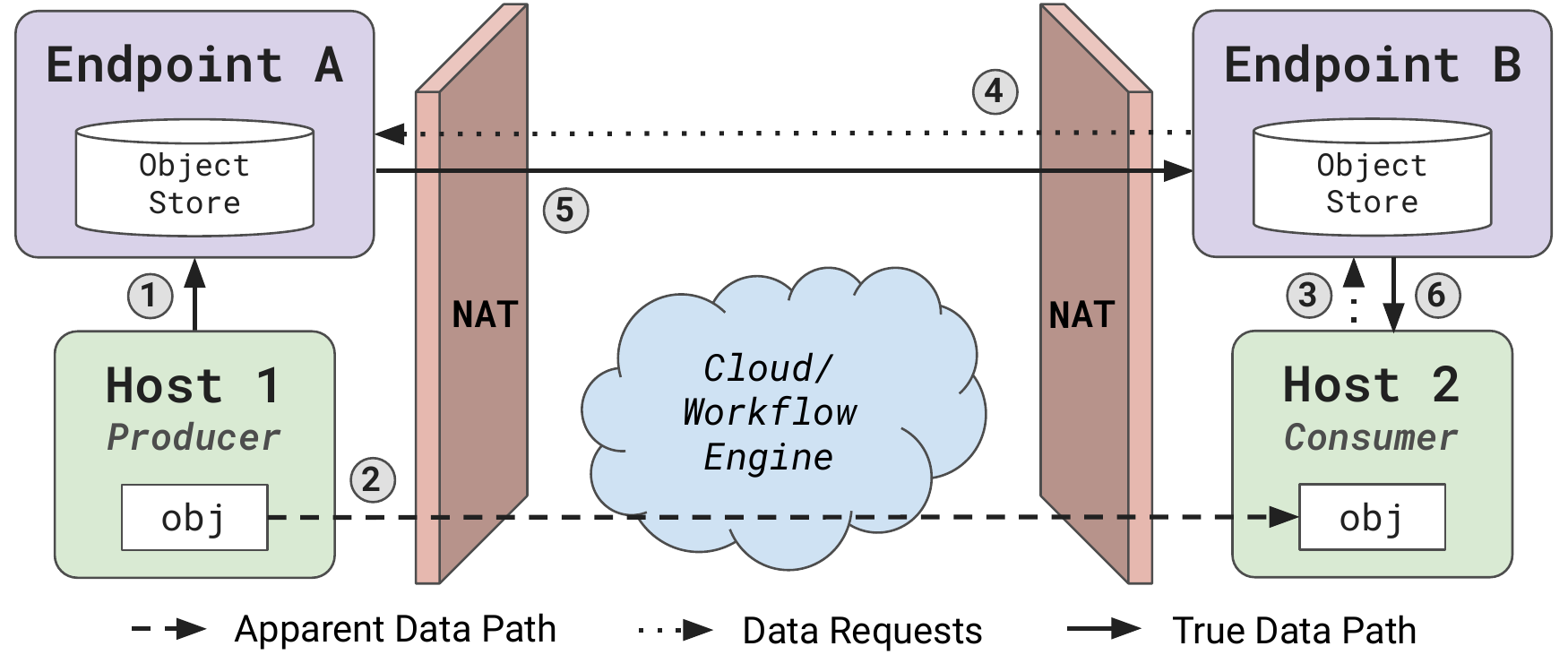}
    \caption{
    Data flow when transferring objects via proxies and \psendpoints between sites.
    The proxy gives the appearance that data flows through the entire application, but the actual data transfer is performed via a peer connection between the \psendpoints at the producing and consuming sites.
    }
    \label{fig:endpoint-overview}
\end{figure}

\begin{figure*}[ht]
    \centering
    \includegraphics[width=\textwidth]{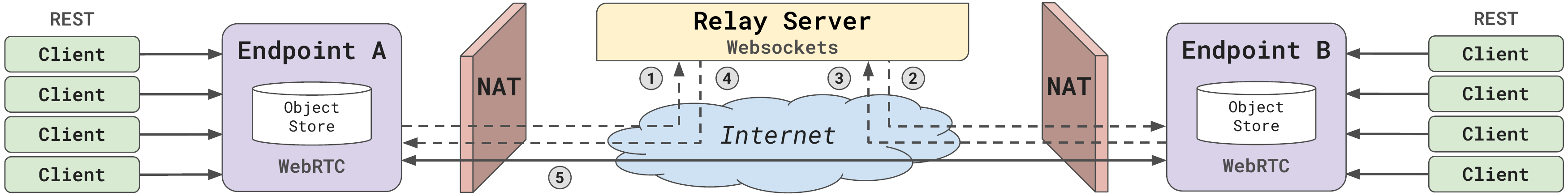}
    \caption{
    Client requests directed to any \psendpoint are forwarded to the correct \psendpoint via a peer connection.
    The peer connections are opened by using UDP hole-punching and a publicly accessible relay server.
    When \psendpoint \emph{A} wants to connect to \psendpoint \emph{B}, \emph{A} asks the relay server \emph{R} to forward a session description protocol (SDP)~\cite{handley2006sdp} to \emph{B} (1 and 2).
    This description contains information about how the two peers can connect, such as what protocols they support.
    \emph{B} receives \emph{A}'s session description from \emph{R} and replies with \emph{B}'s session description (3 and 4).
    \emph{A} and \emph{B} then generate interactive connectivity establishment (ICE) candidates~\cite{keranen2018interactive} (i.e., public IPs and ports to try for the connection) which they exchange via \emph{R}.
    Once \emph{A} and \emph{B} have exchanged ICE candidates, they can connect by completing the hole punching process (5).
    }
    \label{fig:endpoint-peering}
\end{figure*}

A common pattern in inter-site applications is the use of a centralized orchestrator that can communicate with all sites and mediates the control flow between actors across the sites.
A simple example is a cloud-hosted queue of tasks, which actors at each site poll to obtain tasks to execute or to place new tasks on the queue.
In this model, data producers may not always know where data are eventually needed, but it can also be prohibitively expensive (monetary or overhead) to store data in the cloud or in some other central service.
The proxy model allows applications to pass data by reference across sites and perform the underlying communication more directly, avoiding additional overheads of unnecessary data movements.
\proxystore{} includes a \proxystore{} endpoint (\psendpoints) model that facilitates direct data transfer between sites as shown in~\autoref{fig:endpoint-overview}.

\psendpoints are in-memory object stores, with optional on-disk storage if host memory is insufficient or data persistence is required.
\psendpoints are managed with the \texttt{\MakeLowercase{\proxystore{}}-endpoint} command-line interface.
Clients use the \texttt{EndpointConnector} to interact with \psendpoints, and object keys are the tuple \texttt{(object\_id, endpoint\_id)}.
If a \psendpoint receives an operation request on a key with an \texttt{endpoint\_id} that is not its own, the \psendpoint establishes a peer connection to the target \psendpoint and forwards the request.

Peer-to-peer communication between \psendpoints is achieved via the Web Real-Time Communication (WebRTC) standard~\cite{WebRTC, blum2021webrtc}---specifically, by using the \emph{RTCPeerConnection} and \emph{RTCDataChannel} components of the \texttt{aiortc} open-source WebRTC implementation~\cite{aiortc}.
The RTCPeerConnection handles the establishment of peer connections across firewalls using NAT traversal and hole punching, as described in~\autoref{sec:related:communication}; security; and connection management.
The RTCDataChannels are associated with an RTCPeerConnection and enable bidirectional transfer between peers;
data are transported over the channel via SCTP (Stream Control Transmission Protocol) over DTLS (Datagram Transport Layer Security).

\psendpoints use a publicly accessible relay server or signaling server to facilitate the creation of RTCPeerConnections.
The process of establishing the connection via the relay server is illustrated in~\autoref{fig:endpoint-peering}.
Once a peer connection is established, the \psendpoints maintain the connection until one of the \psendpoints is stopped; the connection is re-established if lost for any reason, e.g., due to a \psendpoint going offline temporarily.
The hosting requirements for the relay server are minimal because establishing a peer connection only requires the relay server to exchange a few small ($O$(KB)) messages between the peers.
We provide a WebSocket-based~\cite{fette2011websocket} relay server implementation that can be self hosted.

\psendpoints are single-threaded, asyncio applications.
When started, they connect and register with the relay server, and the relay server assigns a unique UUID if not already assigned.
An asyncio task is created which listens on the WebSocket connection with the relay server for incoming peering requests and responds appropriately.
Once a peer connection is established, the \psendpoint also listens for and responds to incoming requests from its peers.

\subsection{The \texttt{MultiConnector} Abstraction}
\label{sec:design:multiconnector}

Sophisticated applications that employ multiple data communication patterns can benefit from using multiple types of mediated communication (i.e., \texttt{Connector} implementations).
Rather than creating multiple \texttt{Store} instances and a policy directing when to use each instance, \proxystore{} provides the \texttt{MultiConnector} abstraction, which is initialized with a mapping of \texttt{Connector} instances to policies, to indicate how each \texttt{Connector} should be used.
Thus, an application can use a single \texttt{Store} instance, and operations will be routed transparently and automatically to the appropriate \texttt{Connector}.
Policy definitions are flexible and can be extended by developers.
An example policy may include minimum and maximum object sizes, representing the ideal operating range for that \texttt{Connector}; tags denoting the sites at which the \texttt{Connector} is accessible (e.g., a \texttt{MargoConnector} is available within a  single cluster, while an \texttt{EndpointConnector} is available at multiple sites); and a prioritization function for breaking ties when multiple \texttt{Connector} instances are otherwise suitable for a given object.

\texttt{Store} operations accept additional keyword arguments that are passed to the corresponding \texttt{Connector} method.
The \texttt{put} and \texttt{proxy} methods of \texttt{MultiConnector} take a set of optional constraints on the data being stored.
These constraints, as well as other metadata (object type or size, location, etc.), are matched against each policy of each \texttt{Connector} managed by the \texttt{MultiConnector}.
If no match is found then an error is raised, although deployments may often prefer to provide a low priority fallback with no constraints.

\section{Evaluation}
\label{sec:experiments}

We evaluate the component-level performance of \proxystore{}, quantify overhead reductions in compute frameworks when using \proxystore{}, and demonstrate the use of \proxystore{} in three real-world scientific applications.
For brevity, we use the term \texttt{XStore} to mean we are using a \proxystore{} \texttt{Store} initialized with an \texttt{XConnector} for communication.
E.g., \texttt{RedisStore} is a \texttt{Store} initialized with a \texttt{RedisConnector}.

We performed experiments using six machines:
Theta, Polaris, Perlmutter, Frontera, Midway2, and Chameleon Cloud.
Theta and Polaris are at Argonne National Laboratory.
Theta is a \num{4392}-node Intel Knights Landing (KNL) cluster.
The 560-node Polaris has four NVIDIA A100 GPUs per node.
NERSC's Perlmutter cluster has \num{1536} NVIDIA A100 GPU nodes and \num{3072} AMD EPYC CPU nodes.
% Theta is an Intel KNL cluster with \num{4392} compute nodes each with 192 GB of RAM.
% ThetaGPU, the GPU subsystem of Theta, has 24 NVIDIA DGXA100 nodes with two 64-core AMD EPYC CPUs, 1 TB of RAM, and eight 40GB A100 GPUs each.
% Polaris is a 560 node GPU cluster with each node containing a 32-core AMD EPYC CPU, 512 GB of RAM, and four 40GB A100 GPUs.
% NERSC's Perlmutter has 1536 GPU nodes and 3072 CPU nodes.
% The GPU nodes have a single 64-core AMD EPYC CPU, 256 GB of RAM, and four 40GB A100 GPUs, and the CPU nodes have two 64-core AMD EPYC CPUs with 256 GB of RAM per socket.
We use the login nodes of \midway at the University of Chicago and the Texas Advanced Computing Center's (TACC) Frontera cluster as clients to distributed applications running on the aforementioned systems.
Chameleon Cloud~\cite{keahey2020lessons} provides bare-metal compute nodes.

\subsection{\proxystore{} with FaaS}
\label{sec:experiments:faas}

We first evaluate \proxystore{} with the federated FaaS platform \globuscompute~\cite{chard20funcx}, with the goal of quantifying the performance gains that may be achieved with minimal code changes to the producer and no changes to the compute framework.

\begin{figure*}
    \centering
    \includegraphics[width=\textwidth]{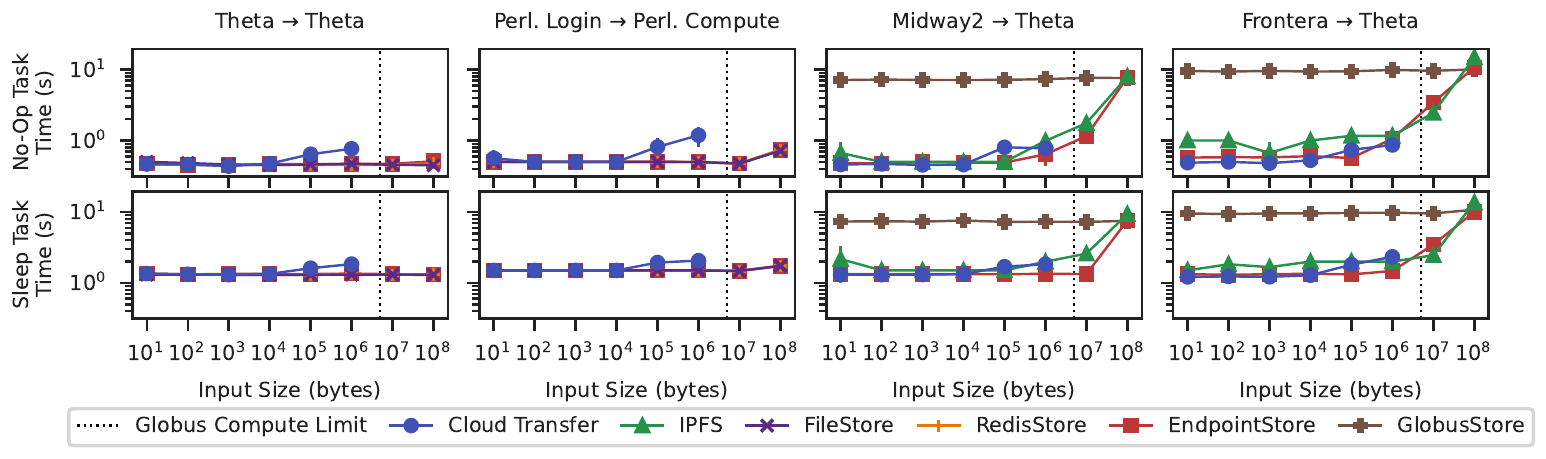}
    \caption{
    Average performance for round-trip \globuscompute no-op (top) and 1~s sleep tasks (bottom), for intra-site (two left columns) and inter-site (two right columns) configurations.
    In intra-site configurations, we compare baseline input data transfer via cloud to \proxystore's \texttt{FileStore}, \texttt{RedisStore}, and \texttt{EndpointStore}.
    For inter-site, we compare to IPFS and \proxystore{}'s \texttt{EndpointStore} and \texttt{GlobusStore}.
    Dashed lines denote the 5~MB \globuscompute payload size limit for transfer via the cloud; \proxystore{} can handle >5~MB task payloads without modifying task code to communicate via alternate means.
    Error bars denote standard deviation.
    }
    \label{fig:globus-compute-roundtrip-task-times}
\end{figure*}

To quantify the extent to which passing task inputs with proxies can reduce data transfer overheads, we perform experiments with \globuscompute where we execute no-op and 1~s sleep tasks with payload sizes from 10 bytes to 100~MB (\autoref{fig:globus-compute-roundtrip-task-times}).
We use four different configurations of \globuscompute clients and endpoints.
(1)~Theta $\rightarrow$ Theta: Client and tasks all run on the same Theta node.
(2)~Perlmutter Login $\rightarrow$ Perlmutter Compute: Client runs on a Perlmutter login node and tasks on a Perlmutter compute node.
(3)~\midway $\rightarrow$ Theta: Client runs on a \midway login node and tasks on a Theta compute node.
(4)~Frontera $\rightarrow$ Theta: Client runs on a Frontera login node and tasks on a Theta compute node.
In the first two scenarios, the client and task execute in the same site and thus we compare the round-trip time when data are moved via the \globuscompute cloud service to data movement via \proxystore's \texttt{FileStore}, \texttt{RedisStore}, and \texttt{EndpointStore}.
In the latter two scenarios, the client and task execute in different sites, so we compare the baseline to \proxystore's \texttt{EndpointStore} and \texttt{GlobusStore}.
We also compare with a configuration in which data are moved to the \globuscompute endpoint by using the InterPlanetary File System (IPFS)~\cite{benet2014ipfs}.
IPFS is a peer-to-peer distributed file system, so we treat the \globuscompute client and \globuscompute endpoint as two nodes of the distributed file system.
In no-op tasks, we ensure that the proxy is resolved even though no computation is performed,
and in the sleep tasks, we begin asynchronously resolving the proxy before sleeping and then wait on the asynchronous resolve after the sleep to simulate overlapping proxy resolution with compute.

\lstset{escapeinside={(*}{*)}}
\begin{lstlisting}[style=PythonStyle, label={lst:proxystore-globus-compute-code}, caption={Example \proxystore{} usage with \globuscompute.}, float, floatplacement=bt]
from globus_compute_sdk import Executor
from (*\MakeLowercase{\proxystore{}}*).connectors.redis import RedisConnector
from (*\MakeLowercase{\proxystore{}}*).store import Store

def my_function(x: MyDataType): ...

store = Store("store-name", RedisConnector(...))(*\label{lst:globus-compute-proxystore-start}*)
data = store.proxy(...)(*\label{lst:globus-compute-proxystore-end}*)

with Executor(...) as gce:
    fut = gce.submit(my_function, data, ...)
    fut.result()
\end{lstlisting}

The baseline round-trip time, where data are transferred along with the task request to the \globuscompute cloud service, increases with data size up to the \globuscompute limit.
In the first two scenarios where the client and task execution occur at the same site, all three \proxystore{} options eliminate the \globuscompute data transfer overhead.
This was achieved with only two client-side lines of code: one to initialize the \texttt{Store} and one to proxy task inputs before submitting to \globuscompute (\autoref{lst:proxystore-globus-compute-code} lines \ref{lst:globus-compute-proxystore-start}--\ref{lst:globus-compute-proxystore-end}).
The asynchronous resolve in the sleep task requires one additional line of code in the task itself, but the overlap of communication and compute can yield benefits.

In the inter-site cases where the clients run on \midway or Frontera login nodes and execute tasks on Theta, we use the \texttt{Globus\-Store} and \texttt{EndpointStore}.
\texttt{GlobusStore} performance is not competitive with the \globuscompute baseline up to \globuscompute's payload limit.
The performance is a consequence of Globus transfer's hybrid software-as-a-service model, which results in high bandwidth for larger transfers but not low latency for small transfers.
However, the benefits of Globus transfer become substantial as data sizes grow beyond those used in this experiment.
The \texttt{EndpointStore} outperforms the baseline, except for no-op tasks between Frontera and Theta where the performance is comparable.
For the largest (100~MB) payloads, \texttt{EndpointStore} performance is less than the theoretical peak of the connection.
We investigate this discrepancy further in~\autoref{sec:experiments:endpoints:peering}.

We also compare to IPFS for inter-site data transfer.
Task data are written to disk, the file is added to IPFS, and the content ID of the IPFS file is passed as input to the \globuscompute task.
When the \globuscompute task is invoked, IPFS is used to retrieve the file, and the data are read back into memory.
Whereas \proxystore{} required two extra client side lines of code, IPFS support required 13 extra lines of code on both the client and task.
The performance of \psendpoints and IPFS for no-op tasks between \midway and Theta are within run-to-run variances of each other.
\psendpoints are faster with the one second sleep tasks because of the asynchronous resolution of proxies.
\psendpoints outperform IPFS for Frontera to Theta transfers due to Frontera having a slower file system and slower transfers between the IPFS peers compared to the \midway $\rightarrow$ Theta scenario.
IPFS and \psendpoints address a different set of problems---IPFS is designed for decentralized and persistent sharing of content-addressable files; however, IPFS has a mature peer-to-peer transfer protocol which we can use as a point of comparison to show that \psendpoints can outperform IPFS.

\begin{figure}
    \centering
    \includegraphics[width=\columnwidth]{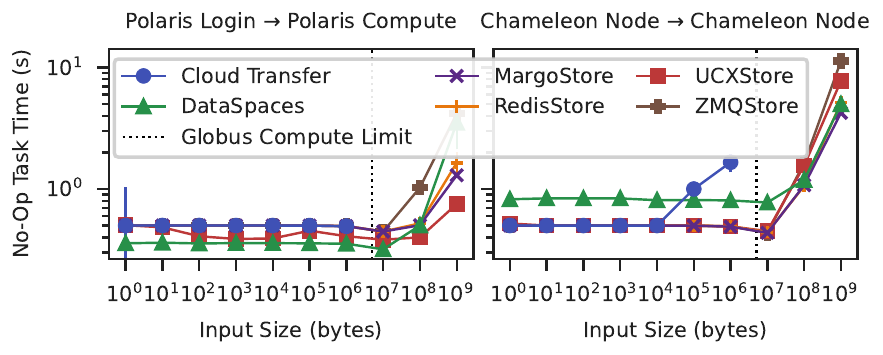}
    \caption{
    Average round-trip performance of no-op \globuscompute tasks on Polaris and Chameleon Cloud for the baseline cloud transfer via the \globuscompute service, \proxystore{} centralized stores
    (\texttt{RedisStore}), \proxystore{} distributed in-memory stores (\texttt{MargoStore},  \texttt{UCXStore}, \texttt{ZMQStore}) and DataSpaces. Error bars denote standard deviation.
    }
    \label{fig:globus-compute-roundtrip-task-times-intrasite}
\end{figure}

We repeat these experiments with the distributed in-memory connectors described in~\autoref{sec:design:intrasite:dim} and compare performance to DataSpaces, a shared-space abstraction designed for large-scale scientific applications. The experiments were executed on Polaris, which has a high-performance HPE Slingshot 11 network, and on two Chameleon Cloud nodes with a Mellanox Connect-X3 40GbE InfiniBand interconnect.
\autoref{fig:globus-compute-roundtrip-task-times-intrasite} shows that the baseline cloud transfer and \proxystore{} alternatives all exhibit similar
performance at data sizes $<$1~GB, after which bandwidth dominates performance.

\texttt{MargoStore} and \texttt{UCXStore}, which both leverage RDMA, achieve the best overall performance on Polaris.
However, \texttt{UCXStore} performs measurably worse than \texttt{MargoStore} and \texttt{RedisStore} for larger data sizes on Chameleon.
We suspect the disparity is a result of the network differences between the two systems.
While we expect DataSpaces and \texttt{MargoStore} to perform similarly because both use Margo for the transport layer, \texttt{MargoStore} outperforms DataSpaces on both systems.
We observed prominent startup overheads, particularly for smaller transfers, with DataSpaces on Chameleon.

We focus on FaaS for HPC and choose FuncX because it is designed to coordinate computation across federated resources (e.g., cloud, HPC, and edge devices). However, \proxystore{} is agnostic to the compute framework and will work with other FaaS systems.
We expect comparable performance characteristics since \globuscompute{}'s data storage and communication mechanisms are similar to cloud-specific FaaS systems.

\subsection{\proxystore{} with Workflow Systems}

\begin{figure}
    \centering
    \includegraphics[width=\columnwidth]{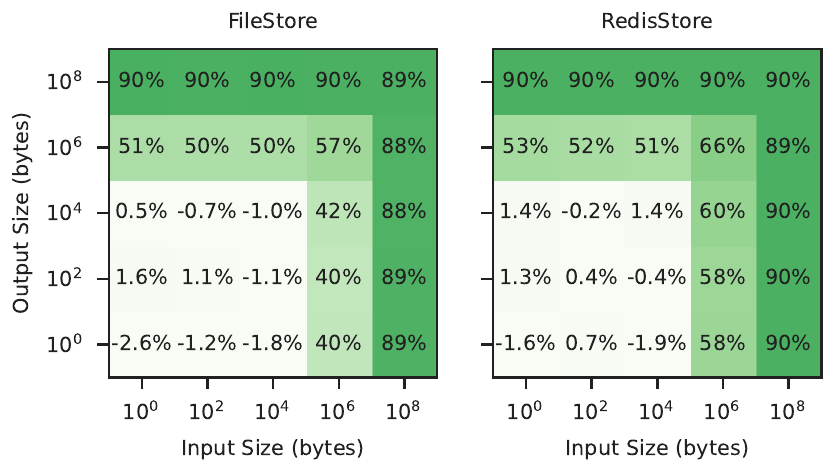}
    \caption{
    Percent improvements in task round-trip time when using \proxystore{} to move data vs.\ Colmena's default method with Parsl.
    Each task configuration is repeated 100 times, and the median time is used to compute the improvement.
    }
    \label{fig:colmena-overheads}
\end{figure}

Colmena is a Python library for steering large ensembles of simulations~\cite{ward2021colmena}.
Colmena applications contain three components: (1) a ``Thinker,'' one or more agents that create tasks and consume results; (2) a ``Task Server,'' which coordinates tasks to be executed by using a workflow engine (here, Parsl); and (3) workers which execute the tasks and return results to the Task Server.
We integrate \proxystore{} into Colmena at the library level.
Users can register a \texttt{Store} and associated threshold for each task type.
Task inputs or results greater than the threshold will be proxied before the task is sent to the Task Server.
Passing proxies with the task can alleviate overheads in the Task Server and underlying workflow system.

We investigate overhead improvements in~\autoref{fig:colmena-overheads}, where we report the percent improvement over the baseline of median round-trip task times.
We execute a series of no-op tasks using Colmena and Parsl with varied input and output sizes.
The Thinker, Task Server, and worker are co-located on a single Theta node to isolate effects of the network.
Neither \proxystore{}'s caching capabilities nor asynchronous resolving of proxies are used.
For small data sizes ($<$100 KB), any improvements in overhead in Colmena are largely negated by the additional overhead of proxying the data (i.e., I/O with storage).
However, \proxystore{} yields 40--60\% improvements in overhead for 1~MB data sizes and 88--89\% for 100~MB data sizes.
This exemplifies why passing by proxy can be invaluable in distributed systems with many interconnected components.
Proxies can be passed around cheaply while ensuring that data are only communicated between producer and consumer.

\subsection{\proxystore{} Endpoint Performance}

To better understand the characteristics of \psendpoints, we next study the times taken for client-to-\psendpoint requests and \psendpoint-to-\psendpoint requests.

\subsubsection{Client Access}

\begin{figure}
    \centering
    \includegraphics[width=\columnwidth]{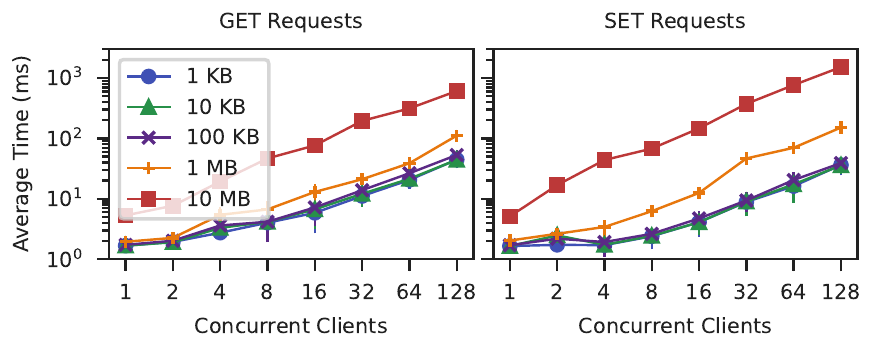}
    \caption{
    Average client \texttt{get} and \texttt{set} request times to a single \psendpoint with respect to payload size and concurrent clients issuing the same request. Error bars denote standard deviation.
    }
    \label{fig:endpoint-clients}
\end{figure}

In~\autoref{fig:endpoint-clients}, we show average per-request times for \texttt{get} and \texttt{set} operations versus the number of concurrent clients making the same request and for varied payload sizes.
Each client makes \num{1000} requests, and the experiment was performed with Python 3.11 on a Perlmutter CPU node.
Response times scale linearly with number of clients for more than two concurrent clients, and also scales with payload size.
This is reasonable given that the proof-of-concept \psendpoint implementation is single-threaded.
Handling many concurrent workers with low latency is better suited for another mediated communication channel such as Redis.

\subsubsection{Endpoint Peering}
\label{sec:experiments:endpoints:peering}

\begin{figure}
    \centering
    \includegraphics[width=\columnwidth]{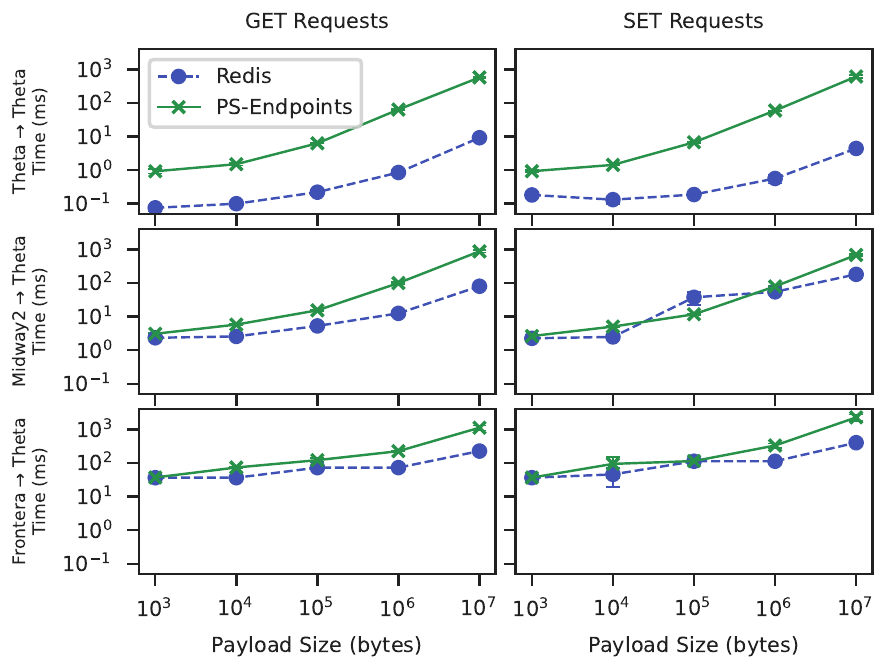}
    \caption{
    Average \texttt{get} and \texttt{set} times, over 1000 requests, between two \psendpoints, with error bars showing the standard deviation.
    Comparisons are made to hosting a Redis server on the target site and opening an SSH tunnel when the two sites are different.
    The \psendpoint{} configuration has one more hop (client---local endpoint---remote endpoint) than Redis (client---remote Redis).
    }
    \label{fig:endpoint-peer-requests}
\end{figure}

The primary use case for \psendpoints is transfers between different sites.
Thus,  we measure request times between \psendpoints as a function of payload size (see \autoref{fig:endpoint-peer-requests}).
We consider three scenarios: requests between two \psendpoints on different Theta nodes, which serves as a baseline; requests between \psendpoints on \midway and Theta; and requests between \psendpoints on Frontera and Theta.
These scenarios differ in latency---packets need only travel tens of meters in the first scenario but \num{1500} kilometers in the third---and bandwidth---the first scenario can utilize the high bandwidth Aries Dragonfly network of Theta while the latter must cross multiple network boundaries.
While no system provides equivalent features to \psendpoints{}, we compare its performance to that of a Redis server hosted on the target site with a (manually created) secure shell (SSH) tunnel between the two sites.
While in practice SSH tunnels can be fragile and difficult to configure (e.g., to  authenticate automatically), they are commonly used by workflow systems~\cite{babuji19parsl, wilde11swift} and the comparison can help highlight strengths and weaknesses of the \psendpoint implementation.

We observe that Redis with SSH is generally faster than \psendpoints, a result for which we identify two primary reasons.
First, the \psendpoint{} configuration has one more hop than the Redis configuration because two endpoints must be used in contrast to a single Redis server and SSH tunnel.
This factor is most prevalent in the Theta-to-Theta scenario where network latency is minimal so the overhead of the extra hop dominates.
Second, we discovered that the \texttt{aiortc} RTCDataChannel cannot fully utilize the available bandwidth between sites.
This is why the difference in performance between \psendpoints and Redis increases at larger data sizes.
A simple test where we established an RTCDataChannel between a process on Frontera and another on Theta achieved a maximum bandwidth of 80~Mbps, a fraction of the full bandwidth available.
This is because computing centers throttle UDP connections to avoid congestion, and \texttt{aiortc} congestion control is slower than other congestion control algorithms like Google's BBR~\cite{googlebbr}.
We support multiplexing data transfer over multiple RTCDataChannels; however, the single-threaded asyncio model is unable to benefit from multiplexing over more than a couple RTCDataChannels.
Despite these networking limitations, the performance of \psendpoints is still competitive with Redis for long distance transfers while not requiring SSH tunnels or open ports.

\subsection{Application: Real Time Defect Analysis}

A common pattern in scientific applications is to transfer data produced by an experiment to a compute facility for analysis.
For example, Argonne National Laboratory's transmission electron microscopy facility uses \globuscompute to invoke a machine-learned segmentation model to quantify radiation damage in acquired images, dispatching this computation to an HPC facility for fast GPU inference.
We modify an open-source real time defect analysis application~\cite{rtdefects} to create and send proxies of images, rather than the actual images. We create a test deployment to mirror the production environment with remotely located instruments and compute.

\begin{table}[t]
\centering
\caption{
Round-trip task times for the real-time defect analysis application.
The \globuscompute endpoint is hosted on a Polaris login node and the tasks are executed on a Polaris compute node.
In the \globuscompute baseline and \texttt{FileStore} configurations, the client (simulating an experimental setup) is hosted on Theta, and the client is hosted on \midway in the \texttt{EndpointStore} configuration.
Transferring task inputs and outputs via \proxystore{} yields >30\% performance improvements in intra- and inter-site task execution.
}
\label{tab:rtdefects}
\begin{center}
\begin{footnotesize}
\begin{tabular}{lccc}
\toprule
\textbf{Configuration}                & \textbf{Proxied}        & \textbf{Time (ms)}   & \textbf{Improvement} \\
\midrule
\globuscompute baseline                         & ---            & $3411 \pm 389$   & ---            \\
\midrule
\multirow{2}{*}{\texttt{FileStore}}     & Inputs         & $2318 \pm 130$   & $32.1\%$          \\
                                        & Inputs/Outputs & $2160 \pm 46$    & $36.6\%$          \\
\midrule
\multirow{2}{*}{\texttt{EndpointStore}} & Inputs         & $2375 \pm 98$    & $30.4\%$           \\
                                        & Inputs/Outputs & $2280 \pm 107$   & $33.2\%$  \\
\bottomrule
\end{tabular}
\end{footnotesize}
\end{center}
\end{table}

We measure the baseline round-trip task time for inference on 1~MB images and compare to \texttt{FileStore} and \texttt{EndpointStore} (\autoref{tab:rtdefects}).
In all cases, we use a \globuscompute endpoint on a Polaris login node that executes tasks on a Polaris compute node.
In the \globuscompute baseline and \texttt{FileStore} cases, our client (i.e., simulated beam facility) is hosted on a Theta login node, and in the \texttt{EndpointStore} case, when the client is on \midway, with \psendpoints on both \midway and a Polaris login node.
We test with only the input images being proxied and with both the input images and inference outputs being proxied.
Note that in the former, the code executed on the \globuscompute endpoint is unchanged, while the latter required two additional lines of task code to proxy the output by using the same \texttt{Store} that was used to resolve the input proxy.

We see in~\autoref{tab:rtdefects} that \proxystore{} improves round-trip task times by 32.1\% and 30.4\% with \texttt{FileStore} and \texttt{EndpointStore}, respectively, when only the inputs are proxied.
Further improvements of a few percentage points can be gained if the downstream code also returns proxies.
We note that \proxystore{} enabled greater flexibility in terms of how clients interact with tasks executed on the \globuscompute endpoint.
Each client can choose its preferred communication method, depending on the mediated communication channels available from itself to the \globuscompute endpoint.

\subsection{Application: Federated Learning}

\begin{figure}[t]
    \centering
    \includegraphics[width=\columnwidth]{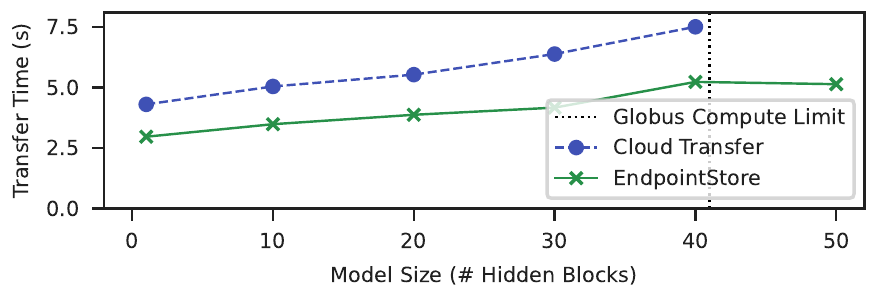}
    \caption{
    Average transfer times for the federated learning use case.
    \psendpoints{} greatly reduce transfer times between nodes compared to cloud transfer.
    In addition, without \proxystore{}, we are unable to transfer models larger than $\sim$40 hidden blocks due to cloud transfer limits.
    }
    \label{fig:flox-result}
\end{figure}

Federated learning~(FL)~\cite{mcmahan2017communication} is an increasingly popular approach to distribute machine learning (ML) training across, often edge~\cite{shi2016edge, mach2017mobile}, devices.
In FL, an aggregator node initializes an ML model and shares it with edge devices to train the model on their own private data in small batches. Once the edge training is complete, the locally-trained models are returned to the aggregator node to ``average'' the model to create a new global model. This new global model is then shared again with the edge devices for further training. In FL only the model is transferred across the network; the distributed edge devices' data are never shared.

Here, we demonstrate the applicability and benefit of \proxystore{} not only for FL use cases but edge computing workflows in general. Due to constrained capacity of edge devices, limited connectivity, and application requirements, making effective use of networks and providing low latency is often crucially important~\cite{mao2017survey}.
\proxystore{} allows for FL control to be separated
from data movement, enabling aggregation to occur \textit{anywhere}, and for models to be transferred directly between edge and aggregation nodes when needed.

Our application is implemented using FLoX~\cite{kotsehub2022flox}, a FL framework which uses \globuscompute to orchestrate training of TensorFlow~\cite{abadi2016tensorflow} models.
Our application trains a convolutional neural network for image classification with the Fashion-MNIST benchmark dataset~\cite{xiao2017/online}.
We increase the number of hidden layers of the neural network to show \proxystore{}'s ability to support larger models compared to a purely FaaS-based approach.
We use the same test bed as used in~\cite{kotsehub2022flox} to deploy our application across four edge devices.
\autoref{fig:flox-result} shows the transfer time as we increase the number of model parameters when using \globuscompute or using \globuscompute and \proxystore{}. We see that \proxystore{} both reduces transfer time and also enables use of larger models.
In the cases where \globuscompute is able to complete the model transfer, \proxystore{} is able to reduce transfer time by $\sim$68\% on average. Further, \proxystore{} can be used to implement hierarchical model aggregation, where sets of edge-trained models are aggregated in a distributed fashion.

\subsection{Application: Molecular Design}

\begin{figure}
    \centering
    \includegraphics[width=\columnwidth]{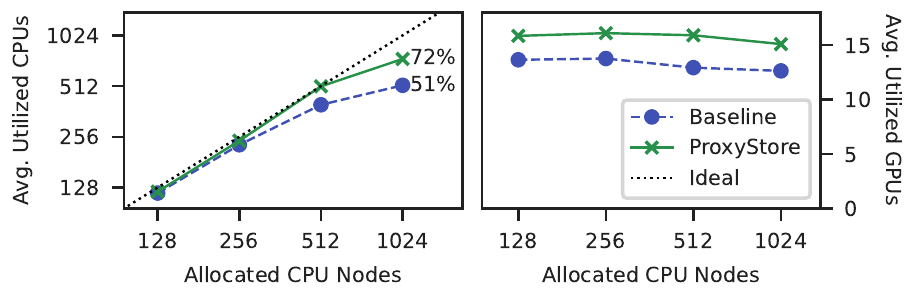}
    \caption{
    Average node utilization of the molecular design application, with and without \proxystore{}.
    The number of GPUs used for training and inference tasks is constant while the number of CPU nodes for simulations is increased.
    Without \proxystore{}, the application struggles to keep all CPU nodes and GPUs fed with tasks.
    \proxystore{} reduces the amount of data flowing through the workflow system, thus reducing the latency between task results being received and new tasks dispatched.
    }
    \label{fig:moldesign-scaling}
\end{figure}

We adapt an open source molecular design workflow to use the \texttt{Multi\-Connector} for communication between tasks.
The workflow uses a mix of quantum chemistry simulations and surrogate machine learning models~\cite{ward2023colmena} to identify electrolytes with high ionization potentials (IP) in a candidate set.

The workflow comprises: (1)~\emph{simulation} tasks that compute IPs on CPUs, (2)~\emph{training} tasks that train surrogate models to predict IPs, and (3)~\emph{inference} tasks that use trained surrogate models to predict IPs, which are then ranked by confidence and used to guide future simulation tasks.
The simulation tasks run on Theta compute nodes, and the training and inference tasks run on a remote GPU node (located behind a different NAT and using a different authentication procedure than Theta).
Tasks are orchestrated with a Colmena Thinker running on a Theta login node and task execution is managed with Parsl.

To optimize communication of task data, we use the \texttt{Multi\-Connector} configured to use \texttt{RedisConnector} for simulation tasks and \texttt{End\-point\-Connector} for training and inference tasks.
\texttt{Redis\-Connector} is suitable for low-latency communication between Theta login and compute nodes and provides persistence when an application spans multiple batch jobs;
\psendpoints enable peer-to-peer transfer of model weights (10~MB in this case) to and from a remote GPU node.
Inputs to inference tasks also require peer-to-peer data transfer to remote GPU nodes.
The inference dataset is static, so while the first round of proxies result in data being moved to the GPU node; proxies for later inference rounds benefit from cached data.
We also investigated using \texttt{GlobusConnector} for data movement. While in this case the dataset was not large enough to benefit from Globus transfers, this would be a good option if a larger dataset were used.
This workflow exemplifies how \proxystore{} can coordinate optimal communication in complex workflows.
We note that no task code needed to be modified to work with the diverse communication methods employed.

In this application, we want to use proxies to reduce overheads in the workflow system.
We evaluate their effectiveness for this purpose by measuring average node utilization during application execution as a function of the number of Theta KNL nodes used for simulations.
We see in~\autoref{fig:moldesign-scaling} that the workflow system struggles to keep nodes fed with new tasks as scale increases.
However, use of \proxystore{} removes data movement burdens from the workflow system and improves scaling, improving utilization by 29\% and 43\% at 512 and 1024 nodes, respectively.
We also observe \proxystore{} improves utilization of the remote GPUs by speeding up data transfer.
At 1024 nodes with \proxystore{}, computation, rather than communication, becomes the bottleneck because simulation results must be processed serially prior to dispatching new simulations.
Processing a simulation result takes $267\pm518$~ms on average in the baseline 1024 node run, but \proxystore{} improves this time by 25\% to $201\pm140$~ms.

\section{Conclusion}
\label{sec:conclusion}

\proxystore{} is a novel framework for facilitating wide-area data management in distributed applications.
The proxy model provides a pass-by-reference-like interface that can work across processes, machines, and sites, and enables data producers to change communication methods dynamically without altering application behavior.
\proxystore{} provides a suite of communication channel implementations intended to meet most requirements and can be extended to other communication methods.
We demonstrated the use of \proxystore{} with FaaS and workflow systems, synthetic benchmarks, and real-world scientific applications.
We showed that \proxystore{} can accelerate a diverse range of distributed applications and enables comparable performance to alternative approaches while avoiding the cumbersome code changes and/or manual deployment and configurations required by alternatives.

In future work, we will investigate support for more communication methods, advanced data management policies for persistence and replication, and wide-area reference counting for object eviction.
It may be useful to allow for data flow semantics on proxies, so that readers of an object block until the object is written, as in Id~\cite{nikhil1989structures}.
We will investigate areas for optimization such as intelligent prefetching and faster peer-to-peer networking protocols.
We plan to explore extension of the proxy model to other languages and other problems in distributed computing (e.g., lazy library loading).
We hope thus to encourage further research in data fabrics for federated applications and enable scientists and engineers to more easily design sophisticated distributed applications.

\section*{Acknowledgments}
This research was supported in part by the Department of Energy (DOE) under Contract DE-AC02-06CH11357, the ExaWorks project and ExaLearn Co-design Center of the Exascale Computing Project (17-SC-20-SC), and the NSF under Grant 2004894.
This research used resources provided by the Argonne Leadership Computing Facility (ALCF), a DOE Office of Science User Facility supported under Contract DE-AC02-06CH11357; the National Energy Research Scientific Computing Center (NERSC), a U.S. Department of Energy Office of Science User Facility located at Lawrence Berkeley National Laboratory, operated under Contract No. DE-AC02-05CH11231; the Texas Advanced Computing Center (TACC) at the University of Texas at Austin; the University of Chicago's Research Computing Center; and the Chameleon testbed supported by the National Science Foundation.

\balance
\bibliographystyle{plain}
\bibliography{proxystore}

\end{document}